\documentclass[sigconf]{acmart}

\usepackage{booktabs} 

\setcopyright{acmcopyright}
\usepackage[english]{babel}
\usepackage{subcaption}
\usepackage[inline]{enumitem}
\usepackage{amsmath}
\usepackage{stmaryrd}
\usepackage{graphicx}
\usepackage{subcaption}
\usepackage{multirow}
\usepackage{kky}

\def\ignore#1{}

\newcommand{\astfootnote}[1]{
\let\oldthefootnote=\thefootnote
\setcounter{footnote}{0}
\renewcommand{\thefootnote}{\fnsymbol{footnote}}
\footnote{#1}
\let\thefootnote=\oldthefootnote
}

\begin{document}
\title{A Hybrid Retrieval-Generation Neural Conversation Model}


%
%

\author{Liu Yang {\footnotemark[1]}$^1$ \quad Junjie Hu$^2$ \quad  Minghui Qiu$^3$ \quad Chen Qu $^1$ \quad  Jianfeng Gao$^4$ \quad W. Bruce Croft$^1$ \quad Xiaodong Liu$^4$ \quad Yelong Shen$^5$ \quad Jingjing Liu$^4$}



\affiliation{%
	\institution{
		$^1$ Center for Intelligent Information Retrieval, University of Massachusetts Amherst  \\
		$^2$ Language Technologies Institute, Carnegie Mellon University   \quad
		$^3$ Alibaba Group \\
		$^4$ Microsoft Research Redmond \quad
		$^5$ Tencent AI Lab}
}
\email{{lyang, chenqu, croft}@cs.umass.edu, junjieh@cs.cmu.edu,minghui.qmh@alibaba-inc.com}
\email{{jfgao, xiaodl,jingjl}@microsoft.com,yelongshen@tencent.com}

\begin{abstract}
	\noindent Intelligent personal assistant systems that are able to have multi-turn conversations with human users are becoming increasingly popular. Most previous research has been focused on using either retrieval-based or generation-based methods to develop such systems. Retrieval-based methods have the advantage of returning fluent and informative responses with great diversity. However, the performance of the methods is limited by the size of the response repository. 
	On the other hand, generation-based methods can produce highly coherent responses on any topics. But the generated responses are often generic and not informative due to the lack of grounding knowledge. In this paper, we propose a hybrid neural conversation model that combines the merits of both response retrieval and generation methods. Experimental results on Twitter and Foursquare data show that the proposed model outperforms both retrieval-based methods and generation-based methods (including a recently proposed knowledge-grounded neural conversation model \cite{DBLP:journals/corr/GhazvininejadBC17}) under both automatic evaluation metrics and human evaluation. We hope that the findings in this study provide new insights on how to integrate text retrieval and text generation models for building conversation systems.
\end{abstract}



%



\copyrightyear{2019} 
\acmYear{2019} 
\acmConference[CIKM '19]{The 28th ACM International Conference on Information and Knowledge Management}{November 3--7, 2019}{Beijing, China}
\acmBooktitle{The 28th ACM International Conference on Information and Knowledge Management (CIKM '19), November 3--7, 2019, Beijing, China}
\acmPrice{15.00}
\acmDOI{10.1145/3357384.3357881}
\acmISBN{978-1-4503-6976-3/19/11}

\fancyhead{}
\settopmatter{printacmref=false, printfolios=false}

\maketitle
\renewcommand*{\thefootnote}{\fnsymbol{footnote}}
\footnotetext[1]{Work primarily done during Liu Yang's internship at Microsoft Research. Liu is at Google currently.}
\renewcommand*{\thefootnote}{\arabic{footnote}}

{\fontsize{8pt}{8pt} \selectfont
	\textbf{ACM Reference Format:}\\
	 Liu Yang, Junjie Hu, Minghui Qiu, Chen Qu, Jianfeng Gao, W. Bruce Croft, Xiaodong Liu, Yelong Shen, Jingjing Liu. 2019. A Hybrid Retrieval-Generation Neural Conversation Model. In \textit{The 28th ACM International Conference on Information and Knowledge Management (CIKM '19), November 3--7, 2019, Beijing, China.} ACM, New York, NY, USA, 10 pages. \url{https://doi.org/10.1145/3357384.3357881}}

\section{Introduction}
\label{sec:intro}



The fast development of artificial intelligence has enabled many intelligent personal assistant systems, such as Amazon Alexa, Apple Siri, Alibaba AliMe, Microsoft Cortana, Google Now and Samsung Bixby.\footnote{For example, over 100M installations of Google Now (Google, \url{http://bit.ly/1wTckVs}); 100M sales of Amazon Alexa devices (TheVerge, \url{https://bit.ly/2FbnzTN}); more than 141M monthly users of Microsoft Cortana (Windowscentral, \url{http://bit.ly/2Dv6TVT}).} As a natural interface for human computer interaction, conversation systems have attracted the attention of researchers in the Information Retrieval (IR), Natural Language Processing (NLP) and Machine Learning (ML) communities, leading to a rapidly growing field referred to as Conversational AI \cite{DBLP:journals/corr/abs-1809-08267}.

Typical task-oriented dialog systems use a  modularized architecture which consists of a natural language understanding module, a dialog state tracker, a dialog policy learning module, and a natural language generation module \cite{Henderson2015MachineLF}. In recent years, fully data-driven end-to-end conversation models have been proposed to reduce hand-crafted features, rules or templates. These methods could be grouped into two different categories: generation-based approaches \cite{DBLP:conf/emnlp/RitterCD11,DBLP:conf/acl/ShangLL15,DBLP:conf/naacl/SordoniGABJMNGD15,DBLP:journals/corr/VinyalsL15,DBLP:conf/emnlp/LiMRJGG16} and retrieval-based approaches \cite{DBLP:journals/corr/JiLL14,DBLP:conf/sigir/YanSW16,DBLP:conf/cikm/YanSZW16,DBLP:conf/sigir/YanZE17,DBLP:conf/sigir/YangQQGZCHC18}. 

Given some conversation context, retrieval-based models try to find the most relevant context-response pairs in a pre-constructed conversational history repository. Some of these methods achieve this in two steps: 1) retrieve a candidate response set with basic retrieval models such as BM25 \cite{Robertson:1994:SEA:188490.188561} or QL \cite{Ponte:1998:LMA:290941.291008}; and 2) re-rank the candidate response set with neural ranking models to find the best matching response \cite{DBLP:conf/sigir/YanSW16,DBLP:conf/cikm/YanSZW16,DBLP:conf/sigir/YanZE17,DBLP:conf/acl/WuWXZL17,DBLP:conf/sigir/YangQQGZCHC18}. These methods can return natural human utterances in the conversational history repository, which is controllable and explainable. Retrieved responses often come with better diversity and richer information compared to generated responses \cite{DBLP:conf/ijcai/SongLNZZY18}. However, the performance of retrieval-based methods is limited by the size of the conversational history repository, especially for long tail contexts that are not covered in the history. Retrieval-based models lack the flexibility of generation-based models, since the set of responses of a retrieval system is fixed once the historical context/response repository is constructed.

\begin{table*}[]
	\small
	\caption{A comparison of retrieval-based methods and generation-based methods for data driven conversation models.} 
	\vspace{-0.4cm}
	\begin{tabular}{ l | p{8cm} | p{6.5cm} }
		\hline \hline
		Item  & Retrieval-based methods                                                   & Generation-based methods                                                          \\ \hline \hline
		Main techniques  & Retrieval models; Neural ranking models                            & Seq2Seq models                                                                    \\ \hline
		Diversity        & Usually good if similar contexts have diverse responses in the repository & Easy to generate bland or universal responses                                                  \\ \hline
		Response length  & Can be very long                                                          & Usually short                                                                     \\ \hline
		Context property & Easy for similar context in the repository; Hard for unseen context & Easy to generalize to unseen context                                              \\ \hline
		Efficiency       & Building index takes long time; Retrieval is fast                  & Training takes long time; Decoding is fast                                \\ \hline
		Flexibility      & Fixed response set once the repository is constructed                  & Can generate new responses not covered in history                   \\ \hline
		Fluency          & Natural human utterances                                                  & Sometimes bad or contain grammar errors                                           \\ \hline
		Bottleneck       & Size and coverage of the repository                                     & Specific responses; Long text;  Sparse data  \\ \hline
		Informativeness  & Easy to retrieve informative content                                                 & Hard to integrate external factual knowledge                                      \\ \hline 
		Controllability  &  Easy to control and explain                                               & Difficult to control the actual generated content \\ \hline \hline
	\end{tabular}
\label{tab:compare_ret_gen_models}
\end{table*}

On the other hand, the generation-based methods could generate highly coherent new responses given the conversation context. Much previous research along this line was based on the Seq2Seq model \cite{DBLP:conf/acl/ShangLL15,DBLP:conf/naacl/SordoniGABJMNGD15,DBLP:journals/corr/VinyalsL15}, where there is an encoder to learn the representation of conversation context as a contextual vector, and a decoder to generate a response sequence conditioning on the contextual vector as well as the generated part of the sequence. The encoder/ decoder could be implemented by an RNN with long short term memory (LSTM) \cite{Hochreiter:1997:LSM:1246443.1246450} hidden units or gated recurrent units (GRU) \cite{DBLP:journals/corr/ChungGCB14}. Although generation-based models can generate new responses for a conversation context, a common problem with generation-based methods is that they are likely to generate very general or universal responses with insufficient information such as ``I don't know'', ``I have no idea'', ``Me too'', ``Yes please''. The generated responses may also contain grammar errors. Ghazvininejad et al. \cite{DBLP:journals/corr/GhazvininejadBC17} proposed a knowledge-grounded neural conversation model in order to infuse the generated responses with more factual information relevant to the conversation context without slot filling. Although they showed that the generated responses from the knowledge-grounded neural conversation model are more informative than the responses from the vanilla Seq2Seq model, their model is still generation-based, and it is not clear how well this model performs compared to retrieval-based methods. A comparison of retrieval-based methods and generation-based methods for end-to-end data driven conversation models is shown in Table \ref{tab:compare_ret_gen_models}. Clearly these two types of methods have their own advantages and disadvantages, it is thus necessary to integrate the merits of these two methods.



To this end, in this paper we study the integration of retrieval-based and generation-based conversation models in an unified framework. The closest prior research to our work is the study on the ensemble of retrieval-based and generation-based conversation models by Song et. al. \cite{DBLP:conf/ijcai/SongLNZZY18}. Their proposed system uses a multi-seq2seq model to generate a response and then adopts a Gradient Boosting Decision Tree (GBDT) ranker to re-rank the generated responses and retrieved responses. However, their method still requires heavy feature engineering to encode the context/ response candidate pairs in order to train the GBDT ranker. They constructed the training data by negative sampling, which may lead to sub-optimal performance, since the sampled negative response candidates could be easily discriminated from the positive response candidates by simple term-matching based features. 


We address these issues by proposing a hybrid neural conversational model with a generation module, a retrieval module and a hybrid ranking module. The generation module generates a response candidate given a conversation context, using a Seq2Seq model consisting of a conversation context encoder, a facts encoder and a response decoder. The retrieval module adopts a ``context-context match'' approach to recall a set of response candidates from the historical context-response repository. The hybrid ranking module is built on the top of neural ranking models to select the best response candidate among retrieved and generated response candidates. The integration of neural ranking models, which can learn representations and matching features for conversation context-response candidate pairs, enables us to minimize feature engineering costs during model development. To construct the training data of the neural ranker for response selection, we propose a distant supervision approach to automatically infer labels for retrieved/ generated response candidates. We evaluate our proposed approach with experiments on Twitter and Foursquare data from a previous work by \citet{DBLP:journals/corr/GhazvininejadBC17}. Experimental results show that the proposed model can outperform both retrieval-based models and generation-based models (including a recently proposed knowledge-grounded neural conversation model \cite{DBLP:journals/corr/GhazvininejadBC17}) on both automatic evaluation and human evaluation.\footnote{Code on Github:  \url{https://github.com/yangliuy/HybridNCM}}








In all, our contributions can be summarized as follows:

 \begin{itemize}
    \item We perform a comparative study of retrieval-based models and generation-based models for the conversational response generation task.
	\item We propose a hybrid neural conversational model to combine response generation and response retrieval with a neural ranking model to reduce feature engineering costs.
	\item For model training, we propose a distant supervision approach to automatically infer labels for retrieved/ generated response candidates. We evaluate the effectiveness of different kinds of distant supervision signals and settings for the hybrid ranking of response candidates.
	\item We run extensive experimental evaluation on retrieval-based, generation-based and hybrid models using the Twitter and Foursquare data. Experimental results show that the proposed hybrid neural conversation model can outperform both retrieval-based and generation-based models on both automatic evaluation and human evaluation. We also perform a qualitative analysis on top responses selected by the neural re-ranker and response generation examples to provide insights. 
\end{itemize}




\section{Related Work}
\label{sec:rel}


\textbf{Retrieval-based Conversation Models.} There have been several recent studies on retrieval based-conversation models \cite{DBLP:conf/acl/WuWXZL17,DBLP:conf/emnlp/ZhouDWZYTLY16,DBLP:conf/sigir/YanSW16,DBLP:conf/cikm/YanSZW16,DBLP:conf/sigir/YanZE17,DBLP:journals/corr/JiLL14,DBLP:journals/corr/LowePSP15,DBLP:journals/corr/YangZZGC17,DBLP:conf/sigir/YangQQGZCHC18,DBLP:conf/acl/QiuYJZHCCL18}. Yan et al. \cite{DBLP:conf/sigir/YanSW16} proposed a retrieval-based conversation system with the deep learning-to-respond schema by concatenating context utterances with the input message as reformulated queries. Wu et al. \cite{DBLP:conf/acl/WuWXZL17} proposed a sequential matching network that matches a response with each utterance in the context on multiple levels of granularity to distill important matching information. Yang et al. \cite{DBLP:conf/sigir/YangQQGZCHC18} considered external knowledge beyond dialog context through pseudo-relevance
feedback and QA correspondence knowledge distillation for multi-turn response ranking. Although retrieval-based methods can return fluent responses with great diversity, these approaches lack the flexibility of generation based methods since the set of responses of a retrieval system is fixed once the historical context/ response repository is constructed in advance. Thus retrieval systems may fail to return any appropriate responses for those unseen conversation context inputs \cite{DBLP:journals/corr/abs-1809-08267}. In our work, we study the integration of retrieval-based and generation-based methods for response generation to combine the merits of these two types of methods.

\textbf{Generation-based Conversation Models.}
There have also been a number of recent studies on conversation response generation with deep learning and reinforcement learning \cite{DBLP:conf/emnlp/RitterCD11,DBLP:conf/acl/ShangLL15,DBLP:conf/naacl/SordoniGABJMNGD15,DBLP:journals/corr/VinyalsL15,DBLP:conf/emnlp/LiMRJGG16,DBLP:conf/acl/LiGBSGD16,DBLP:conf/acl/TianYMSFZ17,bordes2017learning,P17-1045,alime-chat,DBLP:journals/corr/abs-1809-05972,P18-1123,DBLP:journals/corr/abs-1806-07042,DBLP:conf/acl/ChengXGLZF18}. \citet{DBLP:journals/corr/abs-1809-08267} performed a comprehensive survey of neural conversation models in this area. Shang et al. \cite{DBLP:conf/acl/ShangLL15} proposed the Neural Responding Machine (NRM), which is an RNN encoder-decoder framework for short text conversations. In order to mitigate the blandness problem of universal responses generated by Seq2Seq models, Li et al. \cite{DBLP:journals/corr/LiGBGD15} proposed the Maximum Mutual Information (MMI) objective function for conversation response generation. Some previous work augments the context encoder to not only represent the conversation history, but also some additional input from external knowledge. Ghazvininejad et al. \cite{DBLP:journals/corr/GhazvininejadBC17} proposed a knowledge-grounded neural conversation model which infuses factual content that is relevant to the conversation context. Qin et al. \cite{qin2019cmr} extended the knowledge-grounded neural conversation model and jointly models response generation and on-demand machine reading, which takes advantage of machine reading models, such as \cite{liu2018san}.   Our research shares a similar motivation with the work by \citet{DBLP:journals/corr/GhazvininejadBC17}, but we do not adopt a pure generation-based approach. Instead, we explore a hybrid approach that combines  retrieval-based models and generation-based models. Similar hybrid approaches are also used in some popular personal intelligent assistant systems including  the ``Core Chat'' component of Microsoft XiaoIce \cite{DBLP:journals/corr/abs-1812-08989}. Our proposed model distinguishes from prior work using the boosted tree ranker~\cite{DBLP:journals/corr/abs-1812-08989,DBLP:conf/ijcai/SongLNZZY18} by using a neural ranking model which holds the advantage of reducing feature engineering efforts for the conversation context/ response candidates pairs during the hybrid re-ranking process.

\textbf{Neural Ranking Models.}
A number of neural ranking models have been proposed for information retrieval, question answering and conversation response ranking \cite{DBLP:conf/cikm/HuangHGDAH13,DBLP:conf/nips/HuLLC14,DBLP:conf/aaai/PangLGXWC16,Guo:2016:DRM:2983323.2983769,Yang:2016:ARS:2983323.2983818,DBLP:conf/acl/WuWXZL17, Xiong:2017:ENA:3077136.3080809,Mitra:2017:LMU:3038912.3052579,alime-tl}. These models could be classified into three categories \cite{Guo:2016:DRM:2983323.2983769,DBLP:journals/corr/abs-1903-06902}. The first category is the \textit{representation-focused} models. These models learn the representations of queries and documents separately and then calculate the similarity score of the learned representations with functions such as cosine, dot, bilinear or tensor layers. A typical example is the DSSM \cite{DBLP:conf/cikm/HuangHGDAH13} model, which is a feed forward neural network with a word hashing phase as the first layer to predict the click probability given a query string and a document title. The second category is the \textit{interaction-focused} models, which build a query-document pairwise interaction matrix to capture the exact matching and semantic matching information between the query-document pairs. The interaction matrix is further fed into deep neural networks which could be a CNN \cite{DBLP:conf/nips/HuLLC14,DBLP:conf/aaai/PangLGXWC16,alime-tl}, term gating network with histogram or value shared weighting mechanism \cite{Guo:2016:DRM:2983323.2983769,Yang:2016:ARS:2983323.2983818} to generate the final ranking score. 
The neural ranking models in the third category combine the ideas of the \textit{representation-focused} models and \textit{interaction-focused} models to jointly learn the lexical matching and semantic matching between queries and documents \cite{Mitra:2017:LMU:3038912.3052579,alime-tl}. The neural ranking model used in our research belongs to the interaction-focused models due to their better performance on a variety of text matching and ranking tasks compared with representation-focused models \cite{DBLP:conf/nips/HuLLC14,DBLP:conf/aaai/PangLGXWC16,Guo:2016:DRM:2983323.2983769,Yang:2016:ARS:2983323.2983818,DBLP:conf/acl/WuWXZL17, Xiong:2017:ENA:3077136.3080809}. 

\section{Our Approach}
\label{sec:our_approach}

\subsection{Problem Formulation}





We define the task of conversational response generation following the previous literature \cite{DBLP:journals/corr/GhazvininejadBC17}. We are given a conversation context $u_i\in\Ucal$, where $u_i$ is the $i$-th context sequence which contains one or multiple utterances. There are also $F$ factual snippets of text $\mathcal{F}_i = \{f_i^1, f_i^2, ..., f_i^F\}$ that are relevant to the $i$-th conversation context $u_i$. Based on the conversation context $u_i$ and the set of external facts $\Fcal_i$, the system outputs an appropriate response which provides useful information to users. Figure 1 shows an example of the conversational response generation task. Given an conversation context ``\textit{Going to Din Tai Fung Dumpling House tonight!}'', we can associate it with several contextually relevant facts from a much larger collection of external knowledge text (e.g., the Wikipedia dump, tips on Foursquare, product customer reviews on Amazon, etc.). A response that is both appropriate and informative in the given example could be ``\textit{The shrimp and pork wontons with spicy sauce are amazing!}''.
 \vspace{-0.2cm}
 
\begin{figure}[th]
	\center
	\includegraphics*[viewport=0mm 0mm 145mm 80mm, scale=0.52]{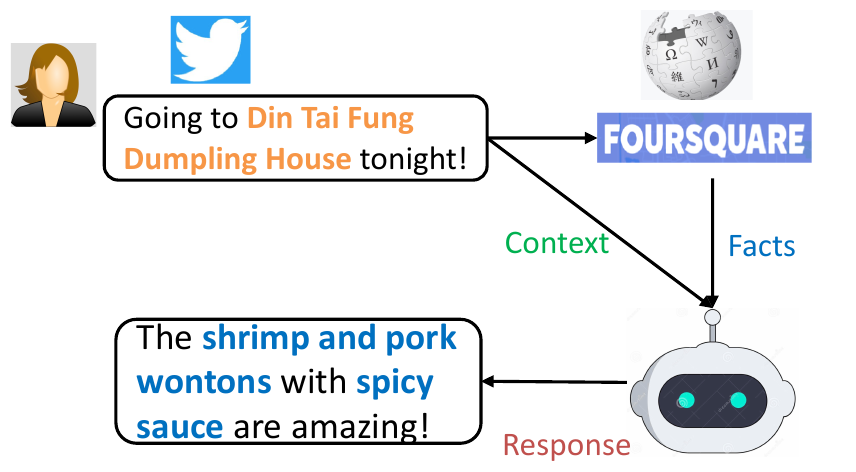}
	\vspace{-0.4cm}
	\caption{An example of the conversational response generation task. The factual information from external knowledge is denoted in blue color.}\label{fig:task-def-resp-gen}
\end{figure}

\begin{figure*}[th]
	\center
	\includegraphics*[viewport=0mm 0mm 280mm 92mm, scale=0.54]{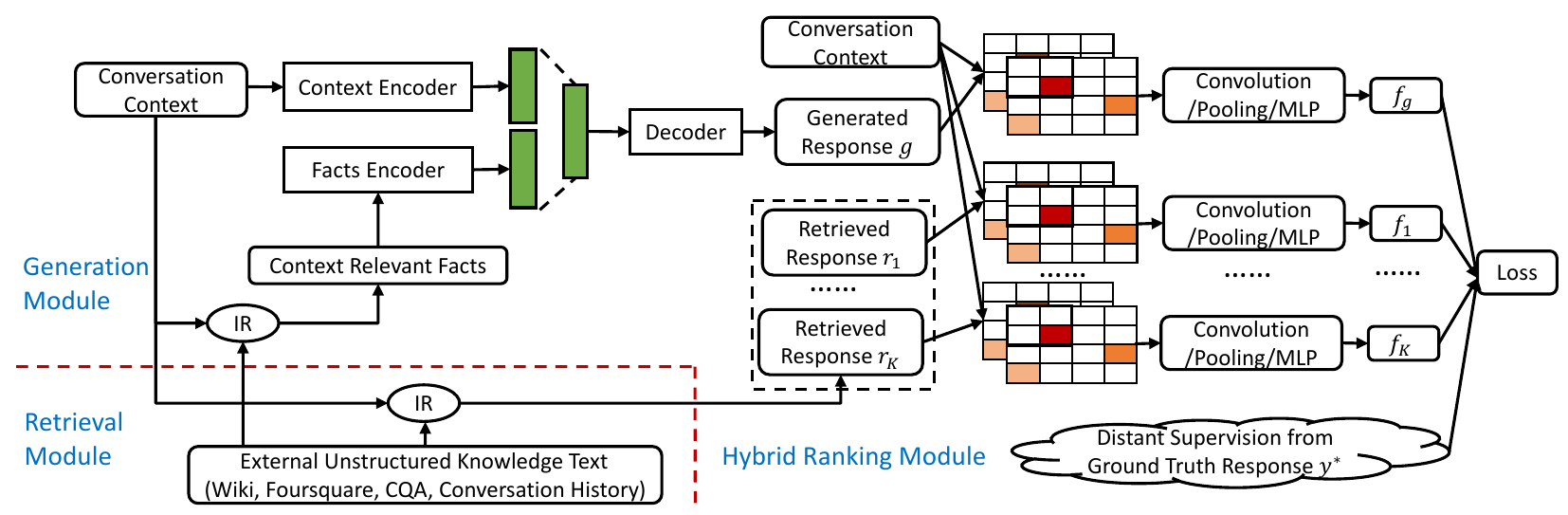}
	\vspace{-0.4cm}
	\caption{The architecture of the Hybrid Neural Conversation Model (HybridNCM). }\label{fig:hncm-model}
\end{figure*}

\begin{table}[!t]
  	\footnotesize
	\caption{A summary of key notations in this work.}
	\vspace{-0.4cm}
	\begin{tabular}
		{ p{0.06\textwidth} | p{0.38\textwidth}} \hline  \hline
		$u_i, \mathcal{U}$ & The context of the $i$-th conversation and the set of all conversation contexts\\\hline
		$f_i^k, \mathcal{F}_i, \mathcal{F}$ & The $k$-th factual text relevant to context $u_i$, the factual texts relevant to context $u_i$ and the set of all factual texts \\ \hline 
		$r_i^k, \mathcal{R}_i$, $\mathcal{R}$ & the $k$-th retrieved response candidate to context $u_i$, the set of all retrieved response candidates for context $u_i$ and the set of all retrieved response candidates\\ \hline
		$g_i^k, \mathcal{G}_i$, $\mathcal{G}$ & the $k$-th generated response candidate to context $u_i$, the set of all generated response candidates for context $u_i$ and the set of all generated response candidates\\ \hline
		$y_i^k, \mathcal{Y}_i$ & the $k$-th response candidate and the union set of all the candidates for the $i$-th context, i.e., $y_i^k \in \Ycal, \Ycal_i=\Rcal_i \cup \Gcal_i$ \\ \hline
		$y_i^*, \mathcal{Y}^*$ & The ground truth response candidate for the i-th context and the set of all ground truth response candidates \\ \hline
		$f(\cdot)$ & The neural ranking model learned in the hybrid ranking module  \\ \hline
		$f( u_i, y_i^k ) $ & The predicted matching score between $u_i$ and $y_i^k$ \\ \hline  \hline 
	\end{tabular}\label{tab:notation}
	\vspace{-10pt}
\end{table}

\vspace{-18pt}
\subsection{Method Overview}
\label{sec:method_overview}


 In the following sections, we describe the proposed Hybrid Neural Conversation Model (HybridNCM) for response generation. Figure \ref{fig:hncm-model} shows the architecture of the hybrid neural conversation model, which consists of three modules:

 	(1) \textbf{Generation Module}: Given the conversation context $u_i $ and the relevant facts $\mathcal{F}_i$, this module is to generate a set of response candidates $\Gcal_i$ using a Seq2Seq model which consists of a conversation context encoder, a facts encoder and a response decoder.
 	
 	(2) \textbf{Retrieval Module}: This module adopts a ``context-context match'' approach to retrieving a few response candidates $\Rcal$. The ``context-context matching'' approach matches the conversation context $u_i $ with all historical conversation context. It then returns the corresponding responses of the top ranked historical conversation context as a set of the retrieved response candidates $\Rcal_i$.
 	
 	(3) \textbf{Hybrid Ranking Module}: Given the generated and retrieved response candidates, i.e., $\Ycal_i=\Gcal_i\cup\Rcal_i$, this module re-ranks all the response candidates with a hybrid neural ranker trained with labels from distant supervision to find the best response as the final system output. 

We will present the details of generating the responses for the $i$-th context $u_i$ by these modules from Section \ref{sec:method_generation_module} to Section \ref{sec:method_hybrid_ranking_module}.  A summary of key notations in this work is presented in Table \ref{tab:notation}. We use a bold letter for a vector or a matrix, and an unbold letter for a word sequence or a set.   




\subsection{Generation Module}\label{sec:method_generation_module}
We map a sequence of words to a sequence of embeddings by looking up the indices in an embedding matrix, e.g., $\ub=\Eb(u_i) = [\ub_1,\ub_2, \cdots, \ub_{L_u}]$ where $L_u$ is the length of a word sequence $u_i$.

\subsubsection{\textbf{Context Encoder}}
\label{sec:context_encoder}
Inspired by previous works on response generation with Seq2Seq models \cite{DBLP:journals/corr/VinyalsL15,DBLP:conf/acl/ShangLL15,DBLP:journals/corr/GhazvininejadBC17}, we adopt a Seq2Seq architecture with attention mechanism \cite{DBLP:journals/corr/BahdanauCB14, DBLP:conf/emnlp/LuongPM15} in the hybrid neural conversation model. In the Seq2Seq architecture, a context encoder is used to transform a sequence of context vectors $\ub = [\ub_1, \ub_2, \cdots, \ub_{L_u}]$ into contextual hidden vectors $\hb = [\hb_1, \hb_2, \cdots, \hb_{L_u}]$ in Eq.~(\ref{Eqn:encoder}). 
	\begin{equation}\label{Eqn:encoder}
	\begin{aligned}
	\hb_t &= \text{RNN}(\ub_t, \hb_{t-1}), 
	\end{aligned}
	\end{equation}
where $\hb_t \in \mathbb{R}^H$ is the hidden state at time step $t$. In our implementation, we stack two layers of LSTM networks as the recurrent neural network. With the context encoder, we can summarize the conversation context by the last hidden vector $\hb_{L_u}$ and maintain the detailed information at each time step by each hidden state $\hb_t$. 

\subsubsection{\textbf{Facts Encoder}}
For the facts encoder, we use the same architecture of the stacked LSTM as the context encoder in Section \ref{sec:context_encoder} to generate the hidden representations of relevant facts. Note that for each conversation context $u_i$, there are $F$ sequences of facts $\Fcal = \{f^1, f^2, \cdots, f^F\}$. We encode these facts into $F$ sequences of hidden vectors $\{\fb^1, \fb^2, \cdots, \fb^F\}$ by the stacked LSTM, where $\fb^j=[\fb^j_1, \fb^j_2, \cdots, \fb^j_L]$ and $L=|\fb^j|$. We summarize a fact into a fixed-size vector by averaging its hidden vectors, i.e., $\bar{\fb}^j=\text{mean}(\fb^j)$. 

\subsubsection{\textbf{Response Decoder}}
The response decoder is trained to predict the next word $g_t$ given the representations of conversation context $\hb_{L_u}$, facts $\bar{\fb}$, and all the previously generated words $g_{1:t-1}$ as follows:
\begin{equation}\label{Eqn:decoder_mle}
\begin{aligned}
p(g | u_i, \Fcal) = \prod_{t=1}^{L_g} p(g_{t} | g_{1:t-1}, u_i, \Fcal)
\end{aligned}
\end{equation}
\begin{align} \label{eq:context_fact}
\Eb &= [\hb_1, \cdots, \hb_{L_u}, \bar{\fb}^1, \cdots, \bar{\fb}^F] \in \mathbb{R}^{H\times (L_u+F)} \\ \label{eq:attention}
\ab_t &= \text{softmax}\rbr{\Eb^T\sbb_{t-1}} \\ \label{eq:attention_context}
\cbb_t &= \Eb \ab_t \\ \label{eq:dec_input} 
\vb_t &= \text{tanh} \rbr{[\sbb_{t-1}, \cbb_t]} \\ \label{eq:decoder}
\sbb_t &= \text{RNN}(\vb_t,\sbb_{t-1}) \\  \label{eq:dec_init}
\sbb_0 &= \varphi \rbr{\text{tanh}\rbr{\hb_{L_u} + {1\over F}\sum_{j=1}^F \bar{\fb}^j}}
\end{align}

For the decoder, we stack two layers of LSTM networks with the attention mechanism proposed in \cite{DBLP:conf/emnlp/LuongPM15}. More specifically, we concatenate the hidden vectors of a context $u_i$ and all factual vectors into a matrix $\Eb$ in Eq.~(\ref{eq:context_fact}). We then compute the attention weight $\ab_t$ by the dot product between the decoder's previous hidden state $\sbb_{t-1}$ and all vectors in $\Eb$, followed by a softmax function in Eq.~(\ref{eq:attention}). The attention context summarizes the conversation context $u_i$ and facts $\Fcal$ by the weighted sum of $\Eb$ in Eq.~(\ref{eq:attention_context}). For the input to the decoder's RNN network, we concatenate the attention context $\cbb_t$ and the previous hidden state $\sbb_{t-1}$ that summarizes the partial generated response $g_{1:t-1}$, and apply a tanh function afterwards in Eq.~(\ref{eq:dec_input}). The initial hidden vector of the decoder is initialized by the last hidden state of the context encoder and the average factual vectors in Eq.~(\ref{eq:dec_init}). $\varphi(\cdot)$ is a linear function that maps a vector from the encoder's hidden space to the decoder's hidden space. The conditional probability at the $t$-th time step can be computed by a linear function $\phi(\cdot)$, which is a fully connected layer, that maps the decoder's hidden state $\sbb_{t-1}$ to a distributional vector over the vocabulary, and a softmax function in Eq.~(\ref{eq:dec_prob}).

\begin{equation}\label{eq:dec_prob}
\begin{aligned}
p(g_t | g_{1:t-1}, u_i, \Fcal) = \text{softmax}(\phi ([\sbb_{t-1}, \cbb_t] ) )
\end{aligned}
\end{equation}
where $\sbb_t$ is the hidden state of the decoder RNN at time step $t$.







 
 
\subsubsection{\textbf{Train and Decode}}  
 Given the ground-truth response $y^*$ to a conversation context $u_i$ with facts $\Fcal$, the training objective is to minimize the negative log-likelihood over all the training data $\mathcal{L}_g$ in Eq.~(\ref{eq:loss_generate}). 
\begin{align} \label{eq:loss_generate}
    \mathcal{L}_g = - {1\over |\Ucal|} \sum_{y^*, u_i, \Fcal} \log p(y^*| u_i, \Fcal)
\end{align}
 During prediction, we use beam search to generate response candidates and perform length normalization by dividing the output log-likelihood score with the length of generated sequences to add penalty on short generated sequences.

\subsection{Retrieval Module}
\label{sec:method_retrieval_module}
The retrieval module retrieves a set of response candidates from the historical conversation context-response repository constructed from the training data. It adopts a ``context-context match'' approach to retrieve a few response candidates. We first index all context/ response pairs in the training data with Lucene.\footnote{\url{http://lucene.apache.org/}} Then for each conversation context $u_i$, we match it with the ``conversation context'' text field in the index with BM25. We return the ``response'' text field of top $K$ ranked context/ response pairs as the retrieved response candidates.\footnote{We set $K=9$ in our experiments.} Note here we only used the context/ response pairs without the facts in the training data. We would like to keep the retrieval module simple and efficient. The re-ranking process of response candidates will be performed in the hybrid ranking module as presented in Section \ref{sec:method_hybrid_ranking_module}.


\subsection{Hybrid Ranking Module}
\label{sec:method_hybrid_ranking_module}
\subsubsection{\textbf{Interaction Matching Matrix}}
We combine a set of generated response candidates $\Gcal_i$ and a set of retrieved response candidates $\Rcal_i$ as the set of all response candidates $\Ycal_i=\Gcal_i\cup\Rcal_i$. 
The hybrid ranking module re-ranks all candidates in $\Ycal_i$ to find the best one as the final system output. In our implementation, $\Gcal_i$ contains one generated response and $\Rcal_i$ contains $K$ retrieved responses.\footnote{We adopt this setting as we find that the generated top responses by Seq2Seq based models are very similar with each other.} Note that facts are not used in this re-ranking process. They are only modeled by the facts encoder in the generation module.  We adopt a neural ranking model following the previous work \cite{DBLP:conf/aaai/PangLGXWC16,DBLP:conf/sigir/YangQQGZCHC18}. Specifically, for each conversation context $u_i$ and response candidate $y_i^k\in \Ycal_i$, we first build an interaction matching matrix. Given $y_i^k$ and $u_{i}$, the model looks up a global embedding dictionary to represent  $y_i^{k}$ and $u_{i}$ as two sequences of embedding vectors $\mathbf{E}(y_i^{k})=[\yb^k_{i,1}, \yb^k_{i,2}, \cdots, \yb^k_{i,L_y}]$ and  $\mathbf{E}(u_{i})=[\ub_{i,1}, \ub_{i,2}, \cdots, \ub_{i,L_u}]$, where $\yb^k_{i,j} \in \mathbb{R}^d$, $\ub_{i,j} \in \mathbb{R}^d$  are the embedding vectors of the $j$-th word in the word sequences $y_i^{k}$ and $u_i$ respectively. The model then builds an interaction matrix $\mathbf{M}$, which computes the pairwise similarity between words in $y_i^{k}$ and $u_i$ via the dot product similarity between the embedding representations. The interaction matching matrix is used as the input of a convolutional neural network (CNN) to learn important matching features, which are aggregated by the final multi-layer perceptron (MLP) to generate a matching score.

\subsubsection{\textbf{CNN Layers and MLP}}
The interaction matrices are fed into a CNN to learn high level matching patterns as features. CNN alternates convolution and max-pooling operations over these inputs. Let $\mathbf{z}^{(l,k)}$ denote the output feature map of the $l$-th layer and $k$-th kernel, the model performs convolution operations and max-pooling operations respectively in Eq.~(\ref{Eqn:2d_conv}) and (\ref{Eqn:2d_max_pooling}).

\textbf{Convolution.} Let $r_w^{(l,k)} \times r_h^{(l,k)}$  denote the shape of the $k$-th convolution kernel in the $l$-th layer, the convolution operation can be defined as:
\begin{footnotesize}
	\begin{equation} \label{Eqn:2d_conv}
	\begin{aligned}
	\mathbf{z}_{i,j}^{(l+1, k)}=\sigma \rbr{\sum_{k'=0}^{K_l -1}\sum_{s=0}^{r_w^{(l,k)} - 1}\sum_{t=0}^{r_h^{(l,k)} - 1} \mathbf{w}_{s,t}^{(l+1,k)} \cdot \mathbf{z}_{i+s, j+t}^{(l,k')} + b^{(l+1,k)}  }  \quad  \\ 
	\forall l =0,2,4,6,\cdots,  
	\end{aligned}
	\end{equation}
\end{footnotesize}
where $\sigma$ is the activation function ReLU, and $\mathbf{w}_{s,t}^{(l+1,k)}$ and $b^{(l+1,k)}$ are the parameters of the $k$-th kernel on the $(l+1)$-th layer to be learned.  $K_l$ is the number of kernels on the $l$-th layer. 

\textbf{Max Pooling.} Let $p_w^{(l,k)} \times p_h^{(l,k)}$  denote the shape of the $k$-th pooling kernel in the $l$-th layer, the max pooling operation can be defined as:
\begin{footnotesize}
	\begin{equation} \label{Eqn:2d_max_pooling}
	\begin{aligned}
	\mathbf{z}_{i,j}^{(l+1,k)} = \max_{0\le s < p_w^{l+1, k}} \max_{0\le t < p_h^{l+1, k}} \mathbf{z}_{i+s,j+t}^{(l,k)} \quad \forall  l = 1,3,5,7, \cdots, 
	\end{aligned}
	\end{equation}
\end{footnotesize}

Finally we feed the output feature representation vectors learned by CNN into a multi-layer perceptron (MLP) to calculate the final matching score $f( u_i, y_i^{k} ) $.

\subsubsection{\textbf{Distant Supervision for Model Training}}
\label{sec:method_distant_supervision}

For model training, we consider a pairwise ranking learning setting. The training data consists of triples $(u_i, y_i^{k+}, y_i^{k-})$, where $y_i^{k+}$ and $y_i^{k-}$ denote the positive and the negative response candidate  for dialog context $u_i$. A challenging problem here is that there is no ground truth ranking labels for all the candidate responses (either the generated response or the retrieved responses) in $\Ycal_i$ given a conversation context $u_i$. The costs for annotating all context-response candidates pairs for model training would be very high. Thus, we generate training data to train the hybrid ranking module with distant supervision inspired by previous work on relation extraction \cite{DBLP:conf/acl/MintzBSJ09}. Specifically we construct $\Ycal_i$ by mixing $K$ retrieved response candidates $\{r_i^1, r_i^2,...,r_i^K\}$ and one generated response candidate  $\{g_i^1\}$. We then score these $K+1$ response candidates with metrics like BLEU/ ROUGE-L by comparing them with the  ground truth responses in the training data of the generation module. Note that in our setting there can be two different types of ground truth: the one for the generation module to train the Seq2Seq models which we have, and the one for the generated/ retrieved response candidates to train the hybrid ranking module which does not exist in the data. Inspired by the way on deriving the supervision signals for relation extraction from Freebase by \citet{DBLP:conf/acl/MintzBSJ09}, here we derive the supervision signals for the hybrid neural ranker from the observed context/ response pairs in the training data of the generation module. Finally we treat the top $k'$ response candidates\footnote{We set $k'=3$ in our experiments.} ranked by BLEU/ ROUGE-L as positive candidates and other responses as negative candidates. In this way, the training labels of response candidates can be inferred by distant supervision.\footnote{Note that we do not have to do such inference during model testing, since we just need to use the trained ranking model to score response candidates instead of computing training loss during model testing.} We perform experiments to evaluate the effectiveness of different kinds of distant supervision signals. In practice, there could be multiple appropriate and diverse responses for a given conversation context. Ideally, we need multiple reference responses for each conversation context, each for a different and relevant response. We leave generating multiple references for a conversation context for distant supervision to future work. We have to point out that it is difficult to collect the data where each context is paired with comprehensive reference responses. Our proposed method can also be easily adapted to the scenario where we have multiple reference responses for a conversation context. 

Given inferred training labels, we can compute the pairwise ranking-based hinge loss, which is defined as:
\begin{equation}
	\mathcal{L}_h = \sum_{i=1}^{I} \max(0, \epsilon - f( u_i, y_i^{k+} )  +  f( u_i, y_i^{k-} )  ) + \lambda ||\Theta||^2_2
	\end{equation}
where $I$ is the total number of triples in the training data. $\lambda ||\Theta||^2_2$ is the regularization term where $\lambda$  denotes the regularization coefficient. $\epsilon$ denotes the margin in the hinge loss. 

\begin{table}[ht]
	\centering
	\small
	\caption{Statistics of experimental data used in this paper.}
	\vspace{-0.4cm}
	\label{tab:exp_data_stat_train_valid_test}
	\begin{tabular}{l|l|l|l}
		\hline  \hline
		Items                     & Train      & Valid  & Test   \\ \hline  \hline
		\# Context-response pairs & 1,059,370  & 2,067  & 2,066  \\ \hline
		\# Facts                  & 43,111,643 & 79,950 & 79,915 \\ \hline
		Avg \# facts per context  & 40.70      & 38.68  & 38.68  \\ \hline
		Avg \# words per facts    & 17.58      & 17.42  & 17.47  \\ \hline
		Avg \# words per context  & 16.66      & 17.85  & 17.66  \\ \hline
		Avg \# words per response & 11.65      & 15.58  & 15.89  \\ \hline  \hline
	\end{tabular}
\end{table}

\section{Experiments}
\label{sec:exps}
 \subsection{Data Set Description}
 \label{sec:data_desc}

We used the same grounded Twitter conversation data set from the study by Ghazvininejad et. al. \cite{DBLP:journals/corr/GhazvininejadBC17}. The data contains 1 million two-turn Twitter conversations. Foursquare tips\footnote{\url{https://foursquare.com/}}  are used as the fact data, which is relevant to the conversation context in the Twitter data. The Twitter conversations contain entities that tie to Foursquare. Then the conversation data is associated with the fact data by identifying Twitter conversation pairs in which the first turn contained either a handle of the entity name or a hashtag that matched a handle appears in the Foursquare tip data. The validation and test sets (around 4K conversations) are created to contain responses that are informative and useful, in order to evaluate conversation systems on their ability to produce contentful responses. The statistics of data are shown in Table \ref{tab:exp_data_stat_train_valid_test}.  

\subsection{Experimental Setup}

\subsubsection{\textbf{Competing Methods.}}
We consider different types of methods for comparison including retrieval-based, generation-based and hybrid retrieval-generation methods as follows:\footnote{We did not compare with \cite{DBLP:conf/ijcai/SongLNZZY18} since the code of both the state-of-the-practice IR system \cite{DBLP:conf/cikm/YanSZW16} and the multi-seq2seq model, which are the two main components of the proposed ensemble model in \cite{DBLP:conf/ijcai/SongLNZZY18}, is not available. 
}

	 \textbf{Seq2Seq.} This is the standard Seq2Seq model with a conversation context encoder and a response decoder, which is the method proposed in \cite{DBLP:journals/corr/VinyalsL15}.
	 
	 \textbf{Seq2Seq-Facts.} This is the Seq2Seq model with an additional facts encoder, which is the generation module in the proposed hybrid neural conversational model. 
	 
	 \textbf{KNCM-MTask-R.} KNCM-MTask-R is the best setting of the knowledge-grounded neural conversation model proposed in the research by Ghazvininejad et al. \cite{DBLP:journals/corr/GhazvininejadBC17} with multi-task learning. This system is trained with 23 million general Twitter conversation data to learn the conversation structure or backbone and 1 million grounded conversation data with associated facts from Foursquare tips. Since we used the same 1 million grounded Twitter conversation data set from this work, our experimental results are directly comparable with response generation results reported by Ghazvininejad et al. \cite{DBLP:journals/corr/GhazvininejadBC17}.
	 
	 \textbf{Retrieval.} This method uses BM25 model \cite{Robertson:1994:SEA:188490.188561} to match the conversation context with conversation context/ response pairs in the historical conversation repository to find the best pair, which is the retrieval module in the proposed model. 
	
	 \textbf{HybridNCM.} This is the method proposed in this paper. It contains two different variations: 1) \textbf{HybridNCM-RS} is a hybrid method by mixing generated response candidates from Seq2Seq and retrieved response candidates from the retrieval module in HybridNCM; 2) \textbf{HybridNCM-RSF} is a hybrid method by mixing generated response candidates from Seq2Seq-Facts and retrieved response candidates from the retrieval module in HybridNCM. 

\subsubsection{\textbf{Evaluation Methodology}.}
Following previous related work \cite{DBLP:conf/naacl/SordoniGABJMNGD15,DBLP:conf/acl/LiGBSGD16,DBLP:journals/corr/GhazvininejadBC17}, we use BLEU and ROUGE-L for the automatic evaluation of the generated responses. The corpus-level BLEU is known to better correlate with human judgments including conversation response generation \cite{DBLP:journals/corr/GalleyBSJAQMGD15} comparing with sentence-level BLEU. We also report lexical diversity as an automatic measure of informativeness and diversity. The lexical diversity metrics include Distinct-1 and Distinct-2, which are respectively the number of distinct unigrams and bigrams divided by the total number of generated words in the responses. In additional to automatic evaluation, we also perform human evaluation (Section \ref{section:human_eval}) of the generated responses of different systems on the \textit{appropriateness} and \textit{informativeness} following previous work \cite{DBLP:journals/corr/GhazvininejadBC17}. 



\begin{table}[]
	\small
	\centering
	\caption{The hyper-parameter settings in the generation-based baselines and the generation module in the proposed hybrid neural conversation model. These settings are the optimized settings tuned with the validation data.} 
	\vspace{-0.4cm}
	\label{tab:hyper_params}
	\begin{tabular}{l|l|l}
		\hline \hline
		Models & Seq2Seq & Seq2Seq-Facts \\ \hline \hline
		Embedding size     & 512     & 256           \\ \hline 
		\# LSTM layers in encoder/decoder     & 2       & 2             \\ \hline
		LSTM hidden state size     & 512     & 256           \\ \hline
		Learning rate    & 0.0001  & 0.001         \\ \hline
		Learning rate decay    & 0.5     & 0.5           \\ \hline
		\# Steps between validation     & 10000   & 5000          \\ \hline
		Patience of early stopping     & 10      & 10            \\ \hline
		Dropout     & 0.3     & 0.3           \\ \hline \hline
	\end{tabular}
\end{table}

\subsubsection{\textbf{ Parameter Settings}} 
All models are implemented with PyTorch\footnote{\url{https://pytorch.org/}} and MatchZoo\footnote{\url{https://github.com/NTMC-Community/MatchZoo}} toolkit. Hyper-parameters are tuned with the validation data. The hyper-parameters in the generation-based baselines and the generation module in the proposed hybrid neural conversation model are shown in Table \ref{tab:hyper_params}. For the hyper-parameter settings in the hybrid ranking module, we set the window size of the convolution and pooling kernels as $(6,6)$. The number of convolution kernels is $64$. The dropout rate is set to $0.5$. The margin in the pairwise-ranking hinge loss is $1.0$. The distant supervision signals and the number of positive samples per context in the hybrid ranking module are tuned with validation data. The used distant supervision signal is BLEU-1 and we treat top $3$ response candidates ranked by BLEU-1 as positive samples. All models are trained on a single Nvidia Titan X GPU by stochastic gradient descent with Adam~\cite{DBLP:journals/corr/KingmaB14} algorithm. The initial learning rate is $0.0001$. The parameters of Adam, $\beta_1$ and $\beta_2$, are $0.9$ and $0.999$ respectively. The batch size is $500$. The maximum conversation context/ response length is $30$. Word embeddings in the neural ranking model will be initialized by the pre-trained GloVe\footnote{https://nlp.stanford.edu/projects/glove/} word vectors and updated during the training process. 





\subsection{Evaluation Results}

\subsubsection{\textbf{Automatic Evaluation}}


We present evaluation results over different methods on Twitter/ Foursquare data in Table \ref{tab:exp_similarity_based_eval}. We summarize our observations as follows: (1) If we compare retrieval-based methods and HybridNCM with pure generation based methods such as Seq2Seq, Seq2Seq-Facts and KNCM-MTask-R, we find that retrieval-based methods and HybridNCM with a retrieval module achieve better performance in terms of BLEU and ROUGE-L. This verifies the competitive performance of retrieval-based methods for conversation response generation reported in previous related works \cite{DBLP:conf/ijcai/SongLNZZY18}. 
(2) Both HybridNCM-RS and HybridHCM-RSF outperform all the baselines including KNCM-MTask-R with multi-task learning proposed recently by Ghazvininejad et al. \cite{DBLP:journals/corr/GhazvininejadBC17} under BLEU and ROUGE-L. The results demonstrate that combining both retrieved and generated response candidates does help produce better responses in conversation systems. For the two variations of HybridNCM, HybridNCM-RSF achieves better BLEU and worse ROUGE-L. Overall the performances of these two variations of HybridNCM are similar to each other. One possible reason is that, the main gain over baselines comes from the retrieval module and the re-ranking process in hybrid ranking module. So the differences in the generation module do not change the results too much.
(3) For lexical diversity metrics like 1-gram/ 2-gram diversity, generation-based methods are far behind retrieval-based methods and HybridNCM, even for KNCM-MTask-R with external grounded knowledge and multi-task learning. This result shows that the retrieved response candidates are more diverse than the response candidates generated by Seq2Seq models. Researchers have studied Maximum Mutual Information (MMI) object functions \cite{DBLP:journals/corr/LiGBGD15} in neural models in order to generate more diverse responses. It would be interesting to compare MMI models with IR models for conversation response generation. We leave this study to future work. 



\begin{table}[]
	\small
	\centering
	\caption{Comparison of different models over the Twitter/ Foursquare data.
		Numbers in bold font mean the result is the best under the metric corresponding to the column. $\ddagger$ means that the improvement from the model on that metric is statistically significant over all baseline methods with $p < 0.05$ measured by the Student's paired t-test. Note that we can only do significance test for ROUGE-L since the other metrics are corpus-level metrics. The results of KNCM-MTask-R are directly cited from Ghazvininejad et al. \cite{DBLP:journals/corr/GhazvininejadBC17} since we used the same 1 million grounded Twitter conversation data set from this work. Thus we don't have the ROUGE-L result for this baseline method.}  
	\vspace{-0.4cm}
	\label{tab:exp_similarity_based_eval}
	\begin{tabular}{l|l|l|l|l}
		\hline \hline
		Method             & BLEU   & ROUGE-L          & Distinct-1 & Distinct-2 \\ \hline \hline
		Seq2Seq            & 0.5032 & 8.4432           & 2.36\%     & 11.18\%    \\ \hline
		Seq2Seq-Facts      & 0.5904 & 8.8291           & 1.91\%     & 7.85\%     \\ \hline
		KNCM-MTask-R       & 1.0800 & \textbackslash{} & 7.08\%     & 21.90\%    \\ \hline
		Retrieval          & 1.2491 & 8.6302           & \textbf{14.68\%}    & \textbf{58.71\%}    \\ \hline \hline
		HybridNCM-RS  &  1.3450 &	\textbf{10.4078}$^\ddagger$ &	11.30\%	& 47.35\%    \\ \hline
		HybridNCM-RSF      & \textbf{1.3695}	 & 10.3445$^\ddagger$	& 11.10\%	& 46.01\%    \\ \hline \hline
	\end{tabular}
\end{table}


\begin{table*}[]
	\small
	\centering
	\caption{Comparison of different models with human evaluation. $\ddagger$ means that the improvement from the model on that metric is statistically significant over all baseline methods with $p < 0.05$ measured by the Student's paired t-test. The agreement score is evaluated by Fleiss' kappa \cite{fleiss1971mns} which is a statistical measure of inter-rater consistency. Agreement scores are comparable to previous results (0.2-0.5) as reported in \cite{DBLP:conf/acl/ShangLL15,DBLP:conf/ijcai/SongLNZZY18}. Higher scores indicate higher agreement degree. The results of KNCM-MTask-R are not included in this table since the generated responses by KNCM-MTask-R  are not available and the code of the  KNCM-MTask-R is also not available.}  
	\vspace{-0.4cm}
	\label{tab:human_eval_res}
	\begin{tabular}{l|c|c|c|c|l|c|c|c|c|l}
		\hline \hline
		Comparision   & \multicolumn{5}{c|}{Appropriateness}                  & \multicolumn{5}{c}{Informativeness}                  \\ \hline
		Method        & Mean   & Bad(0)  & Neutral(+1) & Good(+2) & Agreement & Mean   & Bad(0)  & Neutral(+1) & Good(+2) & Agreement \\ \hline \hline
		Seq2Seq       & 0.4733 & 61.67\% & 29.33\%     & 9.00\%   & 0.2852    & 0.2417 & 77.58\% & 20.67\%     & 1.75\%   & 0.4731    \\ \hline
		Seq2Seq-Facts & 0.4758 & 62.50\% & 27.42\%     & 10.08\%  & 0.3057    & 0.3142 & 70.75\% & 27.08\%     & 2.17\%   & 0.4946    \\ \hline
		Retrieval     & 0.9425 & 34.42\% & 36.92\%     & 28.67\%  & 0.2664    & 0.8008 & 35.50\% & 48.92\%     & 15.58\%  & 0.3196    \\ \hline \hline
		HybridNCM-RS  & \textbf{1.1175}$^\ddagger$ & 27.83\% & 32.58\%     & 39.58\%  & 0.3010    & \textbf{1.0650}$^\ddagger$ & 18.42\% & 56.67\%     & 24.92\%  & 0.1911    \\ \hline
		HybridNCM-RSF & 1.0358$^\ddagger$ & 31.67\% & 33.08\%     & 35.25\%  & 0.2909    & 1.0292$^\ddagger$ & 20.42\% & 56.25\%     & 23.33\%  & 0.2248    \\ \hline \hline
	\end{tabular}
\end{table*}


\begin{table}[]
	\small
	\centering
	\caption{Side-by-side human evaluation results. Win/Tie/Loss
		are the percentages of conversation contexts a method improves, does not change, or hurts, compared with the method after ``v.s.'' on human evaluation scores. HNCM denotes HybridNCM. Seq2Seq-F denotes Seq2Seq-Facts.}  
	\vspace{-0.4cm}
	\label{tab:human_eval_res_sbs}
	\begin{tabular}{l|l|l}
		\hline \hline
		Type                             & Appropriateness & Informativeness \\ \hline
		Comparision                      & Win/Tie/Loss    & Win/Tie/Loss    \\ \hline \hline
		HNCM-RS~~~v.s. Seq2Seq        & 0.71/0.15/0.14  & 0.84/0.10/0.06  \\ \hline
		HNCM-RSF~v.s. Seq2Seq       & 0.68/0.16/0.16  & 0.82/0.11/0.07  \\ \hline
		HNCM-RS~~~v.s. Seq2Seq-F  & 0.70/0.15/0.15  & 0.80/0.12/0.08  \\ \hline
		HNCM-RSF~v.s. Seq2Seq-F & 0.65/0.19/0.17  & 0.77/0.15/0.09  \\ \hline
		HNCM-RS~~~v.s. Retrieval      & 0.43/0.31/0.26  & 0.50/0.31/0.18  \\ \hline
		HNCM-RSF~v.s. Retrieval     & 0.41/0.30/0.29  & 0.50/0.28/0.22  \\ \hline \hline
	\end{tabular}
\end{table}

\subsubsection{\textbf{Human Evaluation}}
\label{section:human_eval}
Automatic evaluation of response generation is still a challenging problem. To complement the automatic evaluation results, we also perform human evaluation to compare the performance of different methods following previous related works \cite{DBLP:conf/acl/ShangLL15,DBLP:journals/corr/GhazvininejadBC17,DBLP:conf/ijcai/SongLNZZY18}. We ask three educated annotators to do the human evaluation. We randomly sample $400$ conversation contexts from the test data, and instruct the annotators to rate the output responses of different systems.\footnote{We mainly performed human evaluation on our methods and three baselines Seq2Seq, Seq2Seq-Facts and Retrieval. We didn't include KNCM-MTask-R into human evaluation since there is no open source code or official implementation from \cite{DBLP:journals/corr/GhazvininejadBC17}. The results of KNCM-MTask-R in Table \ref{tab:exp_similarity_based_eval} are cited numbers from \cite{DBLP:journals/corr/GhazvininejadBC17} since we used the same experimental data sets.} We hide the system ids and randomly permute the output responses to rule out human bias. In the annotation guidelines, we ask the annotators to evaluate the quality of output responses by different systems from the following 2 dimensions:

\begin{itemize}
	\item \textit{Appropriateness}: evaluate whether the output response is appropriate and relevant to the given conversation context. 
	\item \textit{Informativeness}: evaluate whether the output response can provide useful and factual information for the users.
\end{itemize}

Three different labels ``0'' (bad), ``+1'' (neutral), ``+2'' (good) are used to evaluate the quality of system output responses.  Table \ref{tab:human_eval_res} shows the comparison of different models with human evaluation. The table contains the mean score, ratio of three different categories of labels and the agreement scores among three annotators. The agreement score is evaluated by Fleiss' kappa \cite{fleiss1971mns} which is a statistical measure of inter-rater consistency. Most agreement scores are in the range from 0.2 to 0.5, which can be interpreted as ``fair agreement'' or ``moderate agreement''. \footnote{\url{https://en.wikipedia.org/wiki/Fleiss\%27_kappa}} The annotators have relative higher agreement scores for the informativeness of generation-based methods like Seq2Seq and Seq2Seq-Facts, since these methods are  likely to generate short responses or even responses containing fluency and grammatical problems. 

We summarize our observations on the human evaluation results in Table \ref{tab:human_eval_res} as follows: (1) For the mean scores, we can see both HybridNCM-RS and HybridNCM-RSF achieve higher average rating scores compared with all baselines, in terms of both appropriateness and informativeness. Human evaluation results verify that hybrid models indeed help improve the response generation performances of conversation systems. For baselines, the retrieval-based baseline is stronger than generation-based baselines. For HybridNCM-RS and HybridNCM-RSF, HybridNCM-RS achieves relatively higher average human rating scores with a small gap. (2) For the ratios of different categories of labels, we can see that more than $72\%$ of output responses by HybridNCM-RS ($68\%$ for HybridNCM-RSF) are labeled as ``good (+2)'' or ``neutral (+1)'' for appropriateness, which means that most output responses of hybrid models are semantically relevant to the conversation contexts. Generation-based methods like Seq2Seq and Seq2Seq-Facts perform worse than both the retrieval-based method and hybrid models. The retrieval-based method, although quite simple, achieves much higher ratios for the categories ``good (+2)'' and ``neutral (+1)'' compared with generation-based methods. For informativeness, the hybrid models HybridNCM-RS and HybridNCM-RSF are still the best, beating both generation-based baselines and retrieval-based baselines. These results show that the re-ranking process in the hybrid ranking module trained with distant supervision in hybrid conversation models can further increase the informativeness of results by promoting response candidates with more factual content. (3) For the statistical significance test, both HybridNCM-RS and HybridNCM-RSF outperform all baseline methods with $p < 0.05$ measured by the Student's paired t-test in terms of human evaluation scores. We also show the side-by-side human evaluation results in Table \ref{tab:human_eval_res_sbs}. The results clearly confirm that performances of hybrid models are better than or comparable to the performances of all baselines for most test conversation contexts.

%
 



\subsection{Analysis of Top Responses Selected by Re-ranker}


\begin{table}[]
	\small
	\centering
	\caption{ The number and percentage of top responses selected by the hybrid ranking module from retrieved/ generated response candidates. \#PickedGenRes is the number of selected responses from generated response candidates. \#PickedRetRes is the number of selected responses from retrieved response candidates. \#PickedTop1BM25 is the number of selected responses which is also ranked as top 1 responses by BM25.} 
	\vspace{-0.4cm}
	\label{tab:top_selected_response}
	\begin{tabular}{l|l|l|l|l}
		\hline \hline
		Item             & \multicolumn{2}{l|}{HybridNCM-RS} & \multicolumn{2}{l}{HybridNCM-RSF} \\ \hline \hline
		\#TestQNum       & 2066                   & 100.00\%                   & 2066                      & 100.00\%                      \\ \hline
		\#PickedGenRes   & 179	& 8.66\% &	275	& 13.31\%                       \\ \hline
		\#PickedRetRes   & 1887	& 91.34\%	& 1791	& 86.69\%                     \\ \hline
		\#PickedTop1BM25 & 279 & 13.50\%	& 253	& 12.25\%                     \\ \hline \hline
	\end{tabular}
\end{table}

\begin{table*}[]
	\small
	\centering
	\caption{The response generation performance when we vary the ratios of positive samples in distant supervision.} 
	\vspace{-0.4cm}
	\label{tab:impact_ratios_positive_samples}
	\begin{tabular}{l|l|l|l|l|l|l|l}
		\hline \hline
		& Supervision & \multicolumn{2}{c|}{BLEU-1}        & \multicolumn{2}{c|}{BLEU-2}        & \multicolumn{2}{c}{ROUGE-L}       \\ \hline
		Model                          & \# Positive & BLEU            & ROUGE-L          & BLEU            & ROUGE-L          & BLEU            & ROUGE-L          \\ \hline \hline
		\multirow{3}{*}{HybridNCM-RS}  & k'=1        & 0.9022          & 8.9596           & 0.7547          & 8.8351           & 1.0964          & 8.9234           \\ \cline{2-8} 
		& k'=2        & 1.0649          & 9.7241           & 1.1099          & 9.9168           & 1.1019          & 9.6216           \\ \cline{2-8} 
		& k'=3        & \textbf{1.3450} & \textbf{10.4078} & \textbf{1.1165} & \textbf{10.1584} & \textbf{1.1435} & \textbf{10.0928} \\ \hline
		\multirow{3}{*}{HybridNCM-RSF} & k'=1        & 1.0223          & 9.2996           & \textbf{1.1027} & 9.2453           & 1.0035          & 9.2812           \\ \cline{2-8} 
		& k'=2        & 1.3284          & 9.8637           & 1.0175          & 9.8562           & \textbf{1.0999} & \textbf{9.8061}  \\ \cline{2-8} 
		& k'=3        & \textbf{1.3695} & \textbf{10.3445} & 0.8239          & \textbf{9.8575}  & 0.9838          & 9.7961           \\ \hline \hline
	\end{tabular}
\end{table*}

The number and percentage of top responses selected from retrieved/ generated response candidates by the neural ranking model are shown in Table \ref{tab:top_selected_response}. We summarize our observation as follows: (1) most picked results ($91.34\%$ for HybridNCM-RS and $86.69\%$ for HybridNCM-RSF) are from the retrieved response candidates. This is reasonable because we have multiple retrieved response candidates but only one generated response candidate. In some cases, generated responses are preferred to retrieved responses. (2) Although the percentage of generated responses is not high, this does not mean we can just directly use the results returned by the retrieval method. If we look at the row ``PickedTop1BM25'', we can find that only very few responses ranked as the 1st by BM25 are ranked as the 1st again by HybridNCM. Thus, HybridNCM changed the order of these responses candidates significantly.
In particular, the hybrid ranking module in HybridNCM did the following two tasks: a) re-evaluate and re-rank the previous generated/ retrieved responses to promote the good response; 
b) try to inject some generated responses by Seq2Seq models into retrieved results if possible.  (3) We notice that response candidates generated by Seq2Seq-Facts model are more likely to be picked compared to those generated by Seq2Seq.  When a generated response contains rich factual content, the hybrid ranking module is more likely to pick it, which also helps boost the BLEU metrics.

\subsection{Impact of Distant Supervision Signals}

\begin{table}[]
	\small
	\centering
	\caption{The response generation performance when we vary different distant supervision signals. This table shows the results for the setting ``k'=3'', where there are $3$ positive response candidates for each conversation context. ``SentBLEU'' denotes using sentence-level BLEU scores as distant supervision signals.} 
	\vspace{-0.4cm}
	\label{tab:impact_dist_supervision_singals}
	\begin{tabular}{l|l|l|l|l}
		\hline \hline
		Model       & \multicolumn{2}{c|}{HybridNCM-RS} & \multicolumn{2}{c}{HybridNCM-RSF} \\ \hline
		Supervision & BLEU            & ROUGE-L         & BLEU            & ROUGE-L          \\ \hline \hline
		BLEU-1     & \textbf{1.3450}	& \textbf{10.4078}	& \textbf{1.3695}	& \textbf{10.3445}        \\ \hline
		BLEU-2     & 1.1165	& 10.1584	& 0.8239	& 9.8575         \\ \hline
		ROUGE-L    & 1.1435	& 10.0928	&0.9838	& 9.7961          \\ \hline
		SentBLEU    & 0.8326 &	9.2887 &	1.0631 &	9.6338   \\ \hline \hline
	\end{tabular}
\end{table}

We investigate the impact of different distant supervision signals on the response generation performance in Table \ref{tab:impact_dist_supervision_singals}. We find that distant supervision signals like BLEU-1 are quite effective for training the hybrid ranking module. The sentence-level BLEU is not a good choice for the distant supervision signal. The reason is that the sentence-level BLEU is computed only based on the n-gram precision statistics for a given sentence pair. This score has a larger variance compared with the corpus-level BLEU. Since sentence-level BLEU scores would become very small smoothed values if there are no 4-gram or trigram matches between two sentences, which may happen frequently in short text pairs.

\subsection{Impact of Ratios of Positive Samples}
We further analyze the impact of the ratios of positive/ negative training samples on the response generation performance. Table \ref{tab:impact_ratios_positive_samples} shows the results. The value of $k'$ is the number of positive response candidates for each conversation context when we train the hybrid ranking module. When $k'=1$, we select one positive candidate from the ground truth responses in the training data, which is equivalent to the negative sampling technique. As $k'$ increases, we construct the positive candidates by selecting one positive sample from the ground truth responses and $k'-1$ positive samples from the top ranked candidates by distant supervision. We find that larger $k'$ can improve the response generation performance. This is reasonable since larger $k'$ means the model can observe more positive training samples and positive/ negative response pairs in the pairwise ranking loss minimization process. However, increasing the value of $k'$ also adds risks of introducing noisy positive training data.
Thus, the value of $k'$ is a hyper-parameter, and needs to be tweaked via trial and error.

\begin{table}[]
	\footnotesize
	\centering
	\caption{Examples of output responses by different methods. $r$  means the response is retrieved. $g$ means the response is generated.  Entities marked with [ENTITY] have been anonymized to avoid potentially negative publicity. ``HNCM'' denotes ``HybridNCM''.}
	\vspace{-0.4cm}
	\label{tab:case_study1}
	\begin{tabular}{p{1.5cm}  | p{0.2cm} | p{6.0cm}}
		\hline  \hline
		\multirow{2}{*}{Context} & \multicolumn{2}{p{6.4cm}}{Donated to the [ENTITY] last night and now I have to listen to automated phone calls. It's enough to make me want to cancel. } \\ 
		\hline
		Method      & r/g & System Output Response        \\ \hline
		Ground Truth     & -    & Ask them to put you on their internal dnc list. They will likely respect this, because future calls can get them charged. \\ \hline
		Seq2Seq     & g    & I didn't get it. I didn't.
		\\ \hline
		Seq2Seq-Facts   & g     & I'm sorry to hear that. Please dm us your email address so we can look into this. Thanks!
		\\ \hline
		Retrieval   & r     & It's a known issue in LA county. I just got an email from my dm and tech and it should be good in 30 minutes or so.
		\\ \hline
		HNCM-RS  & r     & We're listening and would like to know more and help with your experience. Please follow us so i can dm you our contact info. [ENTITY] 
		\\ \hline
		HNCM-RSF    & g   & We're sorry to hear this. Please dm us if you need assistance. Please dm us your contact info so we can look into this.
		\\ \hline \hline
	\end{tabular}
\end{table}

\subsection{Examples and Case Study}


We perform a case study in Table \ref{tab:case_study1} on the outputs by different methods. In this example, we can find that the response produced by Seq2Seq is very general and it does not provide any useful information for the user. Seq2Seq-Facts generates a much better response by injecting more factual content into response generation process. The response returned by the Retrieval method is also relevant to the context. However, it provides very specific information like ``LA county'', ``30 minutes'', which may have negative impact on the appropriateness of this response for some users. The responses produced by hybrid models achieve a good balance between specificity and generalization. The response by HybridNCM-RS is from retrieved results and the response by HybridNCM-RSF is from generated results, which shows that both retrieval-based methods and generation-based methods have the capacity to produce good responses for certain contexts. Thus it is a natural to combine these two different types of methods for response generation.




\section{Conclusions and Future Work}
\label{sec:conclu}

In this paper, we perform a comparative study of retrieval-based and generation-based methods for building conversation systems. We  propose a hybrid neural conversation model with the capability of both response retrieval and generation in order to combine the merits of these two types of methods. For the training of the hybrid ranking module, we propose a distant supervision approach to automatically infer labels for retrieved/ generated response candidates. Experimental results with Twitter/ Foursquare data show that the proposed model outperforms both retrieval-based and generation-based methods including a recently proposed knowledge-grounded neural conversation model under both automatic evaluation and human evaluation. The findings in this study provide insights on how to integrate text retrieval and text generation models for building conversation systems. For future work, we would like to study reinforcement learning methods for response selection in order to directly optimize metrics like BLEU/ ROUGE. 



\section{Acknowledgments}
This work was supported in part by the Center for Intelligent Information Retrieval and in part by NSF IIS-1715095. Any opinions, findings and conclusions or recommendations expressed in this material are those of the authors and do not necessarily reflect those of the sponsor. The authors would like to thank Kaixi Zhang for the contribution on data annotation/ proofreading on this work.

\bibliographystyle{ACM-Reference-Format}
\bibliography{reference}

\end{document}